\documentclass[prd,showpacs,preprintnumbers,amsmath,amssymb]{revtex4-1}
\usepackage{hyperref}
\usepackage{graphicx}% Include figure files
\usepackage{dcolumn}% Align table columns on decimal point
\usepackage{bm}% bold math

\def\be{\begin{equation}}
\def\ee{\end{equation}}
\def\bea{\begin{eqnarray}}
\def\eea{\end{eqnarray}}
\def\ba{\begin{array}}
\def\ea{\end{array}}

\begin{document}

\preprint{UTHET-10-1001}

\title{Holographic superconductors near the Breitenlohner-Freedman bound}% Force line breaks with \\

\author{George Siopsis}%
 \email{siopsis@tennessee.edu}
\author{Jason Therrien}%
 \email{jtherrie@utk.edu}
\affiliation{%
Department of Physics and Astronomy,
The University of Tennessee,
Knoxville, TN 37996 - 1200, USA.
}%
\author{Suphot Musiri}
\affiliation{Department of Physics,
Srinakharinwirot University,
Bangkok 10110, Thailand.}
\email{suphot@swu.ac.th}

\date{October 2011}% It is always \today, today,

\begin{abstract}
We discuss holographic superconductors in an arbitrary dimension whose dual black holes have scalar hair of mass near the Breitenlohner-Freedman bound.
We concentrate on low temperatures in the probe limit.
We show analytically that when the bound is saturated, the condensate diverges at low temperatures as $|\ln T|^\delta$, where $\delta$ depends on the dimension.
This mild divergence was missed in earlier numerical studies.
We calculate the conductivity analytically and show that at low temperatures,
all poles move toward the real axis. We obtain an increasingly large number of poles which approach the zeroes of the Airy function in 2+1 dimensions and of the Gamma function in 3+1 dimensions.
Our analytic results are in good agreement with numerical results whenever the latter are available.
\end{abstract}

\maketitle

\section{Introduction}

Using the AdS/CFT correspondence \cite{Maldacena:1997re} it has been shown that if an abelian symmetry has been broken outside of a black hole in AdS space,
scalar hair can form creating a holographic superconductor in the dual CFT\cite{Gubser:2008px,Hartnoll:2008vx,Horowitz:2008bn,Franco:2009yz}. In the last few years it has been seen that these systems exhibit several characteristics seen in real world strongly coupled superconductors and seem very promising. The AdS/CFT
correspondence has also been applied to other areas of condensed matter physics \cite{Hartnoll:2007ai,Hartnoll:2007ih,
Hartnoll:2007ip, Hartnoll:2008hs}; some reviews are \cite{Hartnoll:2009sz,Herzog:2009xv,Horowitz:2010gk}. 

Studying the conductivity of holographic superconductors,
% that has been of great interest is how the system responds to an electromagnetic perturbation. It was first noted by 
Horowitz and Roberts \cite{Horowitz:2008bn,Horowitz:2009ij} noted that at the Breitenlohner-Freedman (BF) bound of the scalar hair of the dual black hole,
quasinormal modes appeared to move toward the real axis at low temperatures.
Once the back reaction to the metric was included, they showed that the quasinormal modes never became normal. Their number was determined by the height of an effective potential associated with the wave equation of an electromagnetic perturbation.
The study was done numerically which limited the ability to go to very low temperatures.

Using the analytic tools developed recently in \cite{Siopsis:2010uq}, we explore the low temperature regime of holographic superconductors near the BF bound.
We find that in the probe limit, when the BF bound is saturated, the condensate diverges at low temperatures as $|\ln T|^\delta$, where $\delta$ is a constant that we compute and depends on the dimension of spacetime. This is a very mild divergence
which explains why it was not detected in earlier numerical studies \cite{Horowitz:2008bn}.
It signals the breakdown of the probe limit at very low temperatures, and by including corrections to the probe limit we determine when this occurs.
When back reaction to the metric is included, the effective potential associated with the electromagnetic perturbation that determines the conductivity has a finite height.
This results in a finite number of quasinormal modes.
As one approaches the probe limit, the height of the potential increases with an attendant increase in the number of modes. The latter approach the real axis as the temperature is lowered at frequencies that we compute analytically. In 2+1 (3+1) dimensions they are given in terms of the zeroes of the Airy (Gamma) function.

We emphasize that, even though we probe the low temperature regime, we do not have access to the zero temperature state. This is because for a given charge of the scalar field, there is a lower bound on the temperature (for scaling dimensions $\Delta \le \frac{d}{2}$) below which the probe approximation breaks down. Thus, even though it is often possible to take a ``zero temperature'' limit of our expressions, the limit itself is {\em unphysical}. It is, however, useful for computational purposes, as {\em physical} low-temperature systems are close to it, as we shall show.

Our paper is organized as follows.
%In section 2 we review the
%holographic superconductor of~\cite{Hartnoll:2008vx} in the probe limit.
In section \ref{sec2} we review the field equations. In section \ref{sec2a} we calculate the critical temperature. In section \ref{sec2b} we discuss the low temperature regime in the probe limit.
We also discuss the validity of the probe limit, and find where it breaks down for a given value of the charge $q$.
In section \ref{sec3} we calculate the conductivity at the BF bound at low temperatures.
Finally, section \ref{sec4} contains
our concluding remarks.

\section{Field equations}
\label{sec2}

We are interested in the dynamics of a scalar field of mass $m$ and electric charge $q$ coupled to a $U(1)$ vector potential in the backgound of a $d + 1 -$~dimensional AdS black hole.
The action is
\be S = \int d^{d+1} \sqrt{-g} \left[ \frac{R + d(d-1)/l^2}{16\pi G} - \frac{1}{4} F_{\mu\nu} F^{\mu\nu} - |(\partial_\mu - iqA_\mu) \Psi|^2 - m^2 |\Psi|^2 \right] \ee
where $F=dA$.
We shall adopt units in which $l=1$.

To find a solution of the field equations, consider the metric {\em ansatz}
\be ds^2 = \frac{1}{z^2} \left[ - f(z) e^{-\chi (z)} dt^2 + d{\vec x}^2 + \frac{dz^2}{f(z)} \right]
\ee
where $\vec x \in \mathbb{R}^{d-1}$, representing an AdS black hole of planar horizon.
The AdS boundary is at $z=0$. We shall choose units so that the horizon is at $z=1$, therefore we require $f(1)=0$.
This is possible because of scaling symmetries of the system and can be done without loss of generality as long as one is careful to only consider physical quantities which are scale invariant.

The Hawking temperature is
\be\label{eqHT} T = - \frac{f'(1)}{4\pi} e^{-\chi (1)/2} \ee
Assuming that the scalar field is a real function $\Psi (z)$ and the potential is an electrostatic scalar potential, $A = \Phi (z) dt$, the
field equations are \cite{Hartnoll:2008vx}
\bea\label{eq1}
\Psi'' + \left[ \frac{f'}{f} - \frac{\chi'}{2} - \frac{d-1}{z} \right] \Psi' + \left[ \frac{q^2\Phi^2 e^\chi }{f^2} - \frac{m^2}{z^2 f} \right] \Psi &=& 0 \nonumber\\
\Phi'' + \left[ \frac{\chi'}{2} - \frac{d-3}{z} \right] \Phi' - \frac{2q^2\Psi^2}{z^2 f} \Phi &=& 0
\nonumber\\
- \frac{d-1}{2} \chi' + z{\Psi'}^2 + \frac{zq^2\Phi^2\Psi^2}{f^2 } e^\chi &=& 0
\nonumber\\
\frac{f}{2} {\Psi'}^2 + \frac{z^2}{4} {\Phi'}^2 e^\chi - \frac{d-1}{2} \frac{f'}{z} + \frac{d(d-1)}{2} \frac{f-1}{z^2} + \frac{m^2\Psi^2}{2z^2} + \frac{q^2\Psi^2 \Phi^2 e^\chi}{2f} &=& 0
\eea
where prime denotes differentiation with respect to $z$,
to be solved in the interval $(0,1)$, where $z=1$ is the horizon and $z=0$ is the boundary.

We are interested in solving the system of non-linear equations (\ref{eq1}) in the limit of large $q$ (probe limit). To this end, we shall expand the fields as series in $1/q$ as follows:
\bea \Psi &=& \frac{1}{q} \left[ \Psi_0 + \Psi_1 \frac{1}{q^2} + \dots \right] \nonumber\\
\Phi &=& \frac{1}{q} \left[ \Phi_0 + \Phi_1 \frac{1}{q^2} + \dots \right] \nonumber\\
f &=& f_0 + f_1 \frac{1}{q^2} + \dots \nonumber\\
\chi &=& \chi_0 + \chi_1 \frac{1}{q^2} + \dots \label{eq1exp}
 \eea
and consider the zeroth order system ($q\to\infty$) first and then discuss the addition of first-order ($\mathcal{O} (1/q^2)$) corrections in order to obtain a physically sensible system.

Near the boundary ($z\to 0$), we have $f\to 1$, $\chi\to 0$ and so approximately
%\be
%z^2 \psi'' - 2z\psi' -m^2 \psi = 0 \ \ , \ \ \ \ \phi'' = 0 \ee
%showing that
\be\label{eq3} \Psi \approx \Psi^{(\pm)} z^{\Delta_\pm} \ \ , \ \ \ \ \Phi \approx \mu - \rho z^{d-2} \ee
where
\be \Delta_\pm = \frac{d}{2} \pm \sqrt{\frac{d^2}{4} + m^2} \ee
While a linear combination of asymptotics is allowed by the field equations, it turns out that any such combination is unstable \cite{Hertog:2004bb}.
However, if the horizon has negative curvature, such linear combinations lead to stable configurations in certain cases \cite{papa1}.

Thus, the system is labeled uniquely by the dimension $\Delta = \Delta_\pm$. The mass of the scalar field is bounded from below by the BF bound \cite{Breitenlohner:1982jf}, $m^2\geq -\frac{d^2}{4}$, and there appears to be a quantum phase transition at $m^2=0$. There is also a unitarity bound that requires $\Delta >\frac{d-2}{2}$.

Demanding at the horizon
\be\label{eq7} \Phi (1) = 0 \ , \ee
(gauge choice ensuring that $A=\Phi dt$ is regular at the horizon \cite{Gubser:2008px}),
$\mu$ is interpreted as the chemical potential of the dual theory on the boundary.
$\rho$ is the charge density on the boundary and the leading coefficient in the expansion of the scalar yields vacuum expectation values of operators of dimension $\Delta_\pm$,
\be\label{eq4} \langle \mathcal{O}_{\Delta_\pm} \rangle = \sqrt 2 \Psi^{(\pm)} \ee
The field equations admit non-vanishing solutions for the scalar below a critical temperature $T_c$ where these operators condense.
% The Hawking temperature is defined after we introduce an arbitrary scale in the system,
% \be T = \frac{d}{4\pi} r_+ \ee
% where $r_+$ is the radius of the horizon in the coordinate system $r = r_+/z$.
% 
In view of (\ref{eq3}) and (\ref{eq4}), it is convenient to define
\be\label{eq8} \Psi (z) = \frac{1}{\sqrt{2} q} b^\Delta z^\Delta F(z) \ \ , \ \ \ \ b = \langle q\mathcal{O}_\Delta \rangle^{1/\Delta} \ee
with $F(0) =1$.

\section{The critical temperature}
\label{sec2a}

Above the critical temperature, $\Psi =0$ and the field equations are solved by the AdS Reissner-N\"ordstrom black hole with flat horizon,
\be\label{eqRN} f(z) = 1 - \left( 1+ \frac{(d-2)\rho^2}{4} \right) z^d + \frac{(d-2)\rho^2}{4} z^{2(d-1)} \ \ , \ \ \ \ \chi(z) = 0 \ \ ,  \ \ \ \ \Phi(z) = \rho \left( 1-z^{d-2} \right) \ee
whose Hawking temperature (\ref{eqHT}) is
\be T = \frac{d}{4\pi} \left[ 1 - \frac{(d-2)^2 \rho^2}{4d} \right] \ee
%The corresponding scale-invariant quantity (reduced critical temperature) is
%\be\label{eq12r} \hat T = \frac{T}{(q\rho)^{1/(d-1)}} \ee
Right at the critical temperature, $\Psi$ obeys the scalar field equation (\ref{eq1}) in the Reissner-N\"ordstrom background (\ref{eqRN}) with $\rho = \rho_c$. Thus $F$ (eq.~(\ref{eq8})) at $T=T_c$ obeys the field equation
\be\label{eq13a} F'' + \left[ \frac{f'}{f} +\frac{2\Delta +1-d}{z} \right] F' + \left[ \Delta \frac{(d- \Delta ) (1-f) + zf'}{z^2 f} + q^2 \rho_c^2 \frac{(1-z^{d-2})^2}{f^2} \right] F = 0\ee
For a given $q$, $\rho_c$ is an eigenvalue which is determined by solving this equation for $F$ subject to the boundary condition at the AdS boundary $F(0)=1$. 
We also demand that at the horizon $F(1)$ be finite, and that
there be no contribution of the other solution (behaving as $F(z)\sim z^{d-2\Delta}$ as $z\to 0$).
The latter condition implies that $F$ has a Taylor expansion around $z=0$ with the properties $F(0)=1$, $F'(0)=0$ (as can be easily deduced from (\ref{eq13a})).

To solve eq.~(\ref{eq13a}) for large $q$ (probe limit), use the expansion (\ref{eq1exp}) to write
\be\label{eq5a} F = F_0 + F_1 \frac{1}{q^2} + \dots \ \ , \ \ \ \
\rho = \frac{1}{q} \left[ \rho_0 + \rho_1 \frac{1}{q^2} + \dots \right]
 \ee
Then at zeroth order ($q\to\infty$ limit), the background (\ref{eqRN}) turns into an AdS Schwarzschild black hole, so
\be f_0(z) = 1 - z^d \ \ , \ \ \ \ \chi_0 = 0 \ee
and we obtain the field equation at the critical temperature
\be\label{eq13Sch} -F_0'' + \frac{1}{z} \left[ \frac{d}{1-z^d} -2\Delta -1 \right] F_0' + \Delta^2 \frac{z^{d-2}}{1-z^d} F_0 = \rho_{0c}^2 \frac{(1-z^{d-2})^2}{(1-z^d)^2} F_0 \ee
%which yields the estimate of
%the reduced (eq.~(\ref{eq12r})) critical temperature
%\be\label{eq11} \hat T_c = \frac{d}{4\pi} \rho_{0c}^{-\frac{1}{d-1}} \ee
The eigenvalue $\rho_{0c}^2$ minimizes the expression
\be\label{eq12a} \rho_{0c}^2 = \frac{\int_0^1 dz \, z^{2\Delta -d+1} \{ (1-z^d) [F_0'(z)]^2 + \Delta^2 z^{d-2} [F_0(z)]^2\} }{\int_0^1 dz \, z^{2\Delta -d+1} \frac{(1-z^{d-2})^2}{1-z^d} [F_0(z)]^2} \ee
We can estimate the eigenvalue by substituting the trial function
\be\label{eq13} F_0 = F_\alpha (z) \equiv 1 - \alpha z^{d-1} \ee
which obeys the boundary conditions $F_\alpha (0)=1$, $F_\alpha'(0) =0$ and $F_\alpha(1)$ is finite.

\begin{figure}[t]
\includegraphics[width=.49\textwidth]{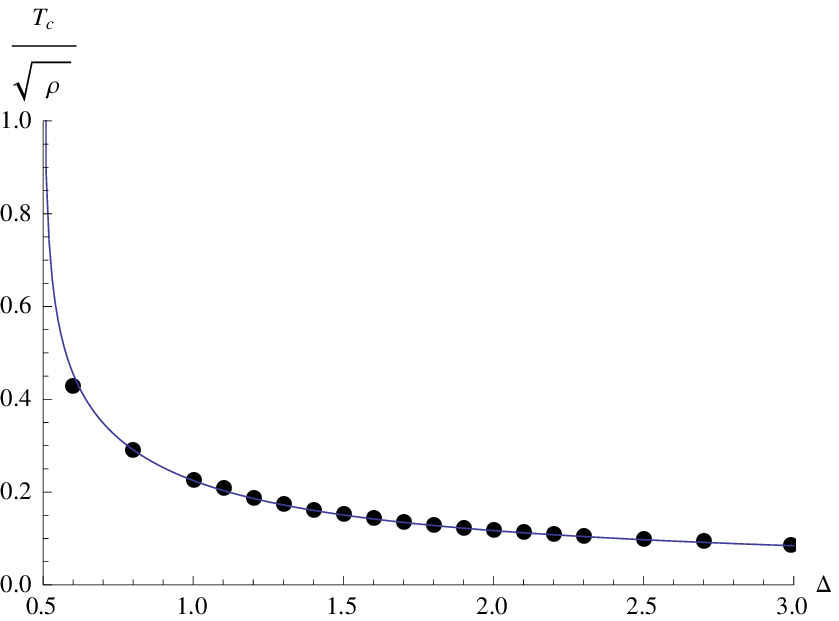}
\includegraphics[width=.49\textwidth]{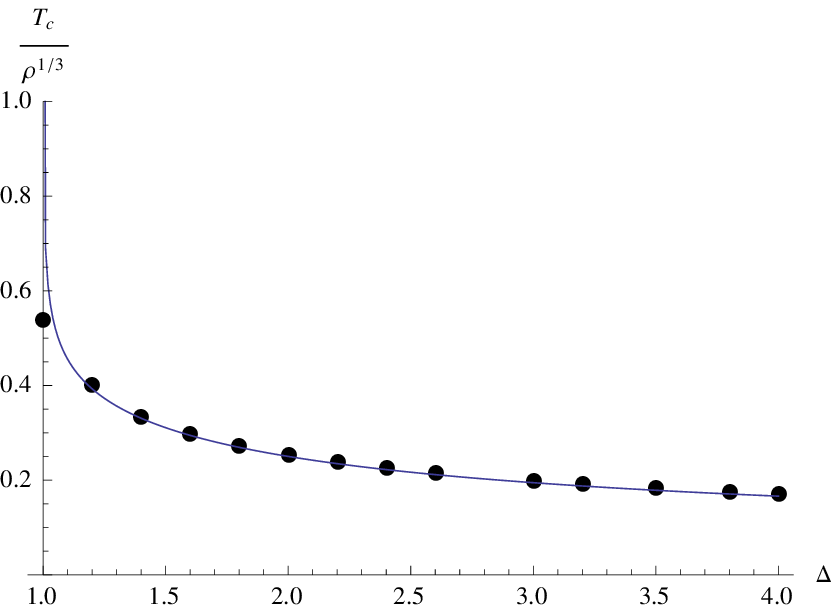}
\caption{The critical temperature $T_c$ {\em vs} the scaling dimension $\Delta$ for $d=3$ (left panel) and $d=4$ (right panel). Data points represent exact values; solid line is obtained by minimizing (\ref{eq12a}) with the trial function (\ref{eq13}).}
\label{fig1}
\end{figure}

For $\Delta = \frac{d}{2}$ and $d=3,4$, we obtain, respectively,
\be \rho_{0c}^2 \approx 6.3, \ 4.2 \ \ , \ \ \ \ T_c \approx 0.15 \sqrt{q\rho} , \ 0.2 (q\rho)^{1/3} \ee
in very good agreement with the exact $T_c = 0.15\sqrt{q\rho},\ 0.25 (q\rho)^{1/3}$.
In fig.~\ref{fig1} we extend the comparison to the entire range of the scaling dimension $\Delta$ for $d=3,4$ demonstrating the accuracy of the estimate (\ref{eq12a}) with the trial function (\ref{eq13}) for the critical temperature.

% Below the critical temperature, the Hawking temperature is found from the last field equation in (\ref{eq1}) to be
% \be T = \frac{1}{4\pi} \left[ d - \frac{1}{4} (\Phi'(1))^2 - \frac{\Delta (d-\Delta) b^{2\Delta}}{4q^2} e^{-\chi (1)} \right] \ee
% showing that at low temperatures, $\rho \sim \mathcal{O} (q^0)$ and $b\sim o (q^{1/\Delta} )$.

\section{Low temperatures}
\label{sec2b}

Next we consider the low temperature regime. Because we are working in units in which the radius of the horizon is fixed ($z=1$), even at low temperatures,
%$\hat T \to 0$ (eq.~(\ref{eq12r})) whereas 
$T$ is bounded. Therefore, $\rho$ becomes large.
Also the condensate diverges in these units for the same reason, and so does $b$ (eq.~(\ref{eq8})).
We are interested in calculating scale-invariant quantities, such as the energy gap
\be\label{eq19} E_g = \frac{\langle q\mathcal{O}\rangle^{1/\Delta} }{ T_c} = \frac{ b}{\rho_0^{1/(d-1)}} + \dots \ee
We shall
consider the probe limit ($q\to\infty$) and then discuss how first-order corrections (in a $1/q$ expansion) can be added to obtain a physically meaningful system.
 
In the probe limit,
as we lower the temperature, $F_0(z)$ (eq.~(\ref{eq5a})) does not have a smooth limit as $b\to \infty$ for $\Delta = \Delta_+ > \frac{d}{2}$.
In this case, care should be exercised in taking this limit which corresponds to the zero temperature limit (ground state) of the system.

We shall concentrate on the case 
$\Delta = \Delta_- \le \frac{d}{2}$ in which $F_0(z)$ has a smooth limit as $b\to \infty$, however the limit system does not correspond to the zero temperature limit and is {\em not} physical. This is because the probe approximation breaks down as $b\to \infty$. As we shall show, the probe expansion is valid for $b \lesssim q^{1/\Delta}$ (eq.\ \eqref{eq50}). So for a given charge $q \gg 1$, we define the low temperature regime by
\be\label{eq21} 1 \lesssim b \lesssim q^{1/\Delta} \ee
%At low temperatures, $b$ is large leading to
%the simplified system of equations (as can be seen by rescaling $z\to z/b$)
We need to solve the system of zeroth-order equations,
\bea\label{eq28} -F_0'' + \frac{1}{z} \left[ \frac{d}{1-z^d}-1-2\Delta \right] F_0' +\frac{\Delta^2 z^{d-2}}{1-z^d} F_0 - \frac{1}{(1-z^d)^2} \Phi_0^2 F_0 &=& 0 \nonumber\\
\Phi_0'' - \frac{d-3}{z} \Phi_0' - \frac{b^{2\Delta} z^{2(\Delta-1)}}{1-z^d} F_0^2 \Phi_0 &=& 0 \eea
Let us first discuss the $b\to \infty$ limit. Even though it is {\em unphysical}, it is well-defined and will be useful for our analytic calculations.

It is evident from the field equation for $\Phi_0$ (and can be easily confirmed numerically for arbitrary regular functions $F_0(z)$) that $\Phi_0 \to 0$ for $z\gtrsim 1/b$ (with $b\gg 1$). Then for $z\gtrsim 1/b$, we obtain
\be F_0(z) = \mathcal{A}\, F \left( \frac{\Delta}{d}, \frac{\Delta}{d}; 1; 1-z^d \right) \ee
which is regular at the horizon. As $b\to \infty$, this becomes valid in the entire interval, because $1/b\to 0$. We deduce the $b\to \infty$ function
\be\label{eqFT0} F_0(z) = \frac{\Gamma^2 (1- \frac{\Delta}{d})}{\Gamma (1-\frac{2\Delta}{d})} \, F \left( \frac{\Delta}{d}, \frac{\Delta}{d}; 1; 1-z^d \right) \ee
where we used $F_0(0)=1$ and standard hypergeometric identities. However, this does not represent the ground state of a physical system, because as $b\to \infty$, the condition for a physical system \eqref{eq21} is violated. Nevertheless, the expression \eqref{eqFT0} is useful for computational purposes.

Next, we wish to solve the equations in the low temperature regime \eqref{eq21}. To this end, we shall use iteration as follows,
\bea\label{eq28i} -{F_0^{(n+1)}}'' + \frac{1}{z} \left[ \frac{d}{1-z^d}-1-2\Delta \right] {F_0^{(n+1)}}' +\frac{\Delta^2 z^{d-2}}{1-z^d} F_0^{(n+1)} &=& \frac{\mu^2}{(1-z^d)^2} [\hat{\Phi}_0^{(n+1)}]^2 F_0^{(n)} \nonumber\\
{\hat{\Phi}_0^{(n+1)\prime\prime}} - \frac{d-3}{z} {{\hat{\Phi}_0^{(n+1)\prime}}} - \frac{b^{2\Delta} z^{2(\Delta-1)}}{1-z^d} [F_0^{(n)}]^2 \hat{\Phi}_0^{(n+1)} &=& 0 \eea
starting with
\be F_0^{(0)} (z) = 1 \ \ , \ \ \ \ \hat{\Phi}_0^{(0)} (z) = 0 \ee
We defined
\be \Phi_0 (z) = \mu \hat\Phi_0 (z) \ \ , \ \ \ \ \hat\Phi_0 (0) = 1 \ee
where $\mu$ is the chemical potential.

At the $n$th step, we obtain the electrostatic potential $\hat{\Phi}_0^{(n+1)}$ from the second equation in \eqref{eq28i} and subsequently the scalar field $F_0^{(n+1)}$ from the first equation in \eqref{eq28i}. To solve the latter, notice that there are three boundary conditions, two at $z=0$ ($F_0^{(n+1)}(0)=1$ and absence of the unwanted alternate behavior $F_0^{(n+1)} \sim z^{d-2\Delta}$), and one at the horizon, $z=1$ (finiteness of $F_0^{(n+1)}$). They are needed to determine the solution as well as the eigenvalue $\mu$.
Using the two boundary conditions at $z=0$, we obtain the solution in terms of the chemical potential,
\bea F_0^{(n+1)} (z) &=& \mathcal{F}_1 (z) \left[ 1 + \mu^2\int_0^z \frac{dz'}{1-(z')^d} (z')^{2\Delta +1 -d} \mathcal{F}_2 (z') [\hat\Phi_0^{(n+1)} (z')]^2 F_0^{(n)}(z') \right]
\nonumber\\
& & - \mathcal{F}_2 (z) \mu^2 \int_0^z \frac{dz'}{1-(z')^d} (z')^{2\Delta +1 -d} \mathcal{F}_1 (z') [\hat\Phi_0^{(n+1)} (z')]^2 F_0^{(n)}(z') \label{eq28n}\eea
where
\be \mathcal{F}_1 (z) = F\left(\frac{\Delta}{d} , \frac{\Delta}{d} ; \frac{2\Delta}{d} ; z^d \right) \ , \ \
\mathcal{F}_2 (z) = \frac{z^{d-2\Delta}}{d-2\Delta} F\left( 1-\frac{\Delta}{d} , 1- \frac{\Delta}{d} ; 2-\frac{2\Delta}{d} ; z^d \right)
\ee
The third boundary condition, regularity at $z=1$, then fixes the chemical potential $\mu$.
%normalization of $\Phi_0^{(n+1)}$.

For $n=0$, we obtain the electrostatic potential
\be\label{eq12phi} \hat\Phi_0^{(1)} (z) = \frac{2}{\Gamma (\nu ) (2\Delta)^\nu} (bz)^{\frac{d-2}{2}} \left[ K_\nu \left( \frac{(bz)^\Delta}{\Delta} \right) - \frac{K_\nu \left( \frac{b^\Delta}{\Delta} \right)}{I_\nu \left( \frac{b^\Delta}{\Delta} \right)} I_\nu \left( \frac{(bz)^\Delta}{\Delta} \right) \right]\ \ , \ \ \ \ \nu = \frac{d-2}{2\Delta} \ee
where we imposed the boundary condition (\ref{eq7}). Notice that the second Bessel function has an exponentially small coefficient, $\mathcal{O} (\sim e^{-2b^\Delta/\Delta} )$, and can be neglected at low temperatures.

The charge density is found by using (\ref{eq3}) and (\ref{eq5a}) to be
\be\label{eq15} \frac{\rho_0}{b^{d-2}} = - \frac{\mu}{ (2\Delta)^{2\nu}} \ee
For the scalar field we obtain eq.\ \eqref{eq28n} with $n=0$,
\be\label{eq12F} F_0^{(1)} (z) = \mathcal{F}_1 (z) \left[ 1 + \mu^2 \int_0^z \frac{dz'}{1-(z')^d} (z')^{2\Delta +1 -d} \mathcal{F}_2 (z') [\hat\Phi_0^{(1)} (z')]^2 \right]
- \mathcal{F}_2 (z) \mu^2\int_0^z \frac{dz'}{1-(z')^d} (z')^{2\Delta +1 -d} \mathcal{F}_1 (z') [\hat\Phi_0^{(1)} (z')]^2 \ee
satisfying the correct boundary conditions at $z=0$. At the horizon ($z=1$), we have a
logarithmic singularity which is found using
\be \mathcal{F}_1 (z) \approx -\frac{\Gamma (\frac{2\Delta}{d})}{\Gamma^2 ( \frac{\Delta}{d} ) } \ln (1-z) \ \ , \ \ \ \ \mathcal{F}_2 (z) \approx -\frac{\Gamma (2-\frac{2\Delta}{d})}{(d-2\Delta)\Gamma^2 (1- \frac{\Delta}{d} ) }\ln (1-z) \ee
Near the horizon, we deduce
\be F_0^{(1)} (z) \approx -\left[ \frac{\Gamma (\frac{2\Delta}{d})}{\Gamma^2 ( \frac{\Delta}{d} ) }(1+ \mu^2 a_2) - \frac{\Gamma (2-\frac{2\Delta}{d})}{(d-2\Delta)\Gamma^2 (1- \frac{\Delta}{d} ) } \mu^2 a_1 \right] \ln (1-z) \ee
where
\be a_i = \int_0^1 \frac{dz}{1-z^d} z^{2\Delta +1 -d} \mathcal{F}_i (z) \left[ \hat\Phi_0^{(1)} (z) \right]^2 \ \ ,  \ \ \ \ i=1,2 \ee
Demanding regularity at the horizon, we need
\be \frac{\Gamma (\frac{2\Delta}{d})}{\Gamma^2 ( \frac{\Delta}{d} ) }(1+ \mu^2 a_2) - \frac{\Gamma (2-\frac{2\Delta}{d})}{(d-2\Delta)\Gamma^2 (1- \frac{\Delta}{d} ) } \mu^2 a_1 = 0 \ee
which fixes the chemical potential $\mu$,
\be\label{eqc1} \frac{1}{\mu^2} = \frac{\Gamma (2-\frac{2\Delta}{d} )\Gamma^2 (\frac{\Delta}{d})}{(d-2\Delta)\Gamma( \frac{2\Delta}{d} )\Gamma^2 ( 1- \frac{\Delta}{d} )} a_1 - a_2 \ee
as advertised.

Explicitly,
%We shall calculate $c$ in the zero temperature limit ($b\to\infty$). To find the leading contribution in this limit, it is convenient to rescale $z\to z/b$. We obtain
\be\label{eqa1a2} a_1 = \frac{1}{b^{2\Delta +2-d}} \frac{(d-2)\Gamma(1-\nu)}{ (2\Delta)^{2\nu}\Gamma (\nu)} + \dots
\ , \ \
a_2 = \frac{1}{b^2} \frac{\sqrt{\pi}\Delta^{\frac{2}{\Delta}-1} \Gamma (\frac{1}{\Delta}) \Gamma (\frac{d-1}{\Delta}) \Gamma (\frac{d}{2\Delta})}{(d-2\Delta)\Gamma^2(\nu) 2^{2\nu}\Gamma ( \frac{d+\Delta}{2\Delta} )} + \dots \ee
Evidently, for $\Delta < \frac{d}{2}$, $a_2\ll a_1$ for $b\ll 1$, therefore
\be\label{eqcb} \mu^2 \approx \mathcal{C} b^{2\Delta +2-d} \ , \ \ \mathcal{C} =  \frac{(d-2\Delta)(2\Delta)^{2\nu}\Gamma(\nu)\Gamma(\frac{2\Delta}{d}) \Gamma^2 (1- \frac{\Delta}{d}) }{(d-2)\Gamma(1-\nu)\Gamma(2-\frac{2\Delta}{d}) \Gamma^2 ( \frac{\Delta}{d}) }
\ee
It is easily seen (using standard hypergeometric identities) that the low temperature expression (\ref{eq12F}) reduces to (\ref{eqFT0}) as $b\to\infty$ in the entire interval $[0,1]$. However, as we have already pointed out, the latter is unphysical. By taking this limit, we do not gain access to a zero temperature system (gound state). Nevertheless, it is useful computationally.

\begin{figure}[t]
\includegraphics[width=.3\textwidth]{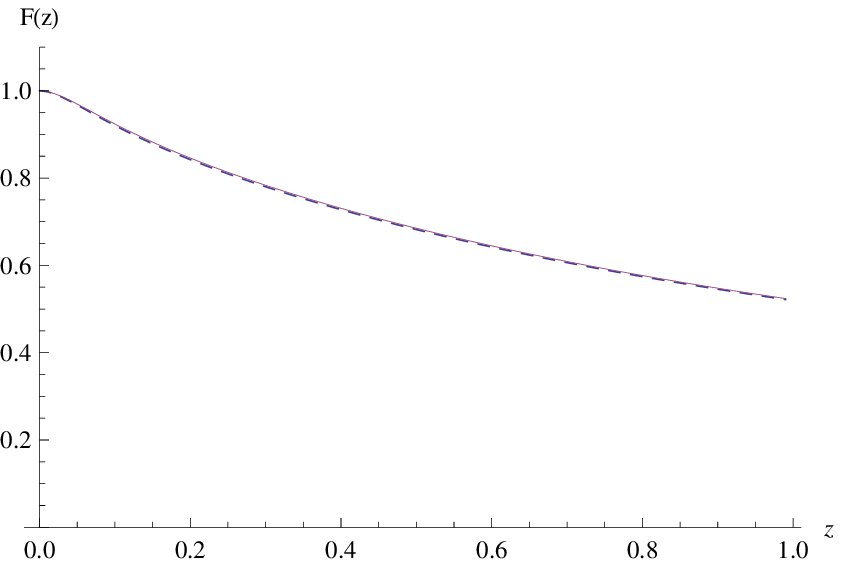}
\includegraphics[width=.3\textwidth]{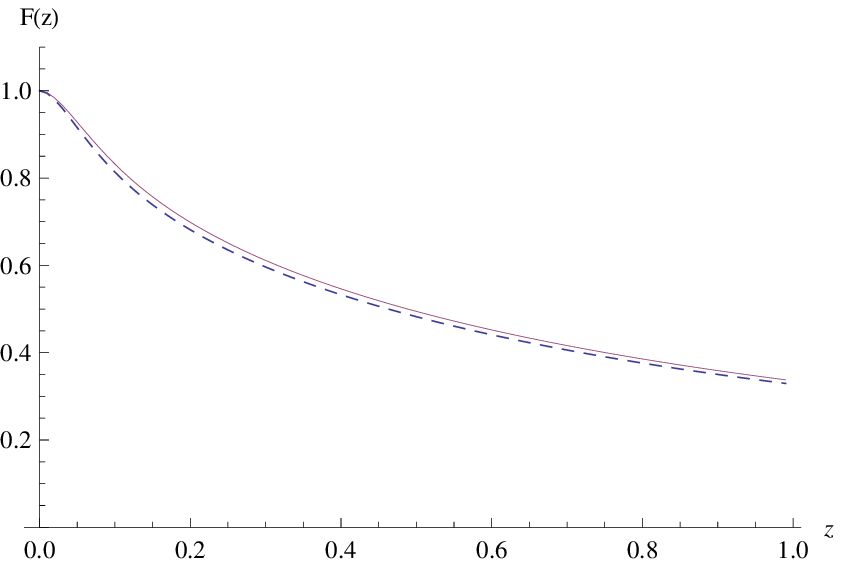}
\includegraphics[width=.3\textwidth]{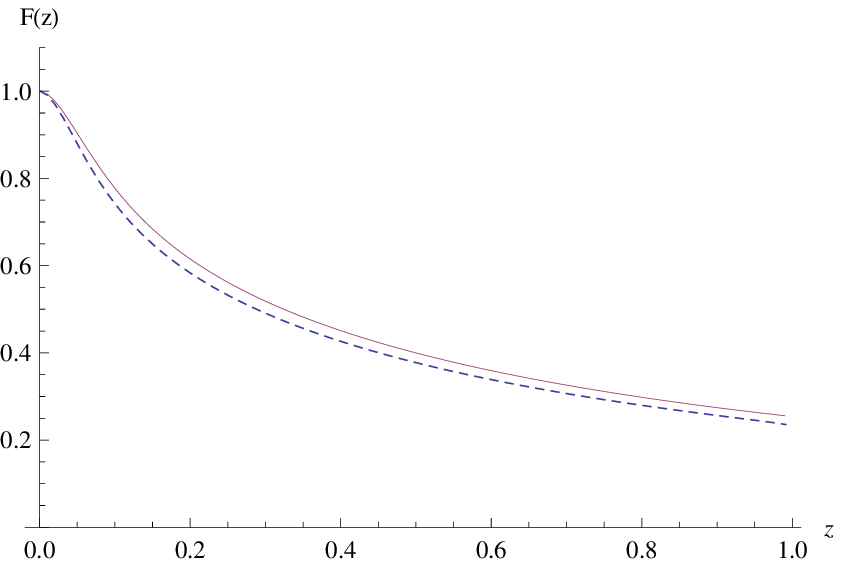}
\caption{The field $F$ (eq.\ (\ref{eq8})) for $\Delta = 1.2$ (left panel), $1.4$ (middle panel), $1.5$ (right panel) and $d=3$. Solid curves are first-order analytic expression (\ref{eq12F}), and dashed curves are exact numerical results (almost indistinguishable) at $T/T_c\approx 0.1$.}
\label{fig2}
\end{figure}
\begin{figure}[ht]
\includegraphics[width=.3\textwidth]{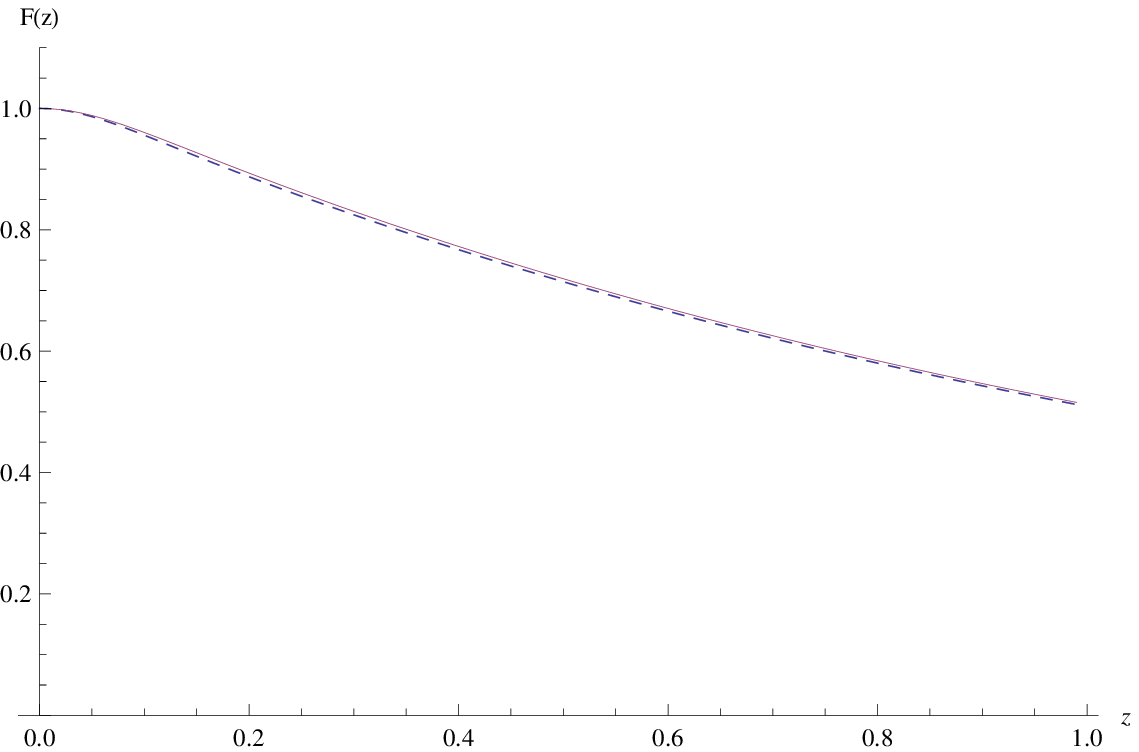}
\includegraphics[width=.3\textwidth]{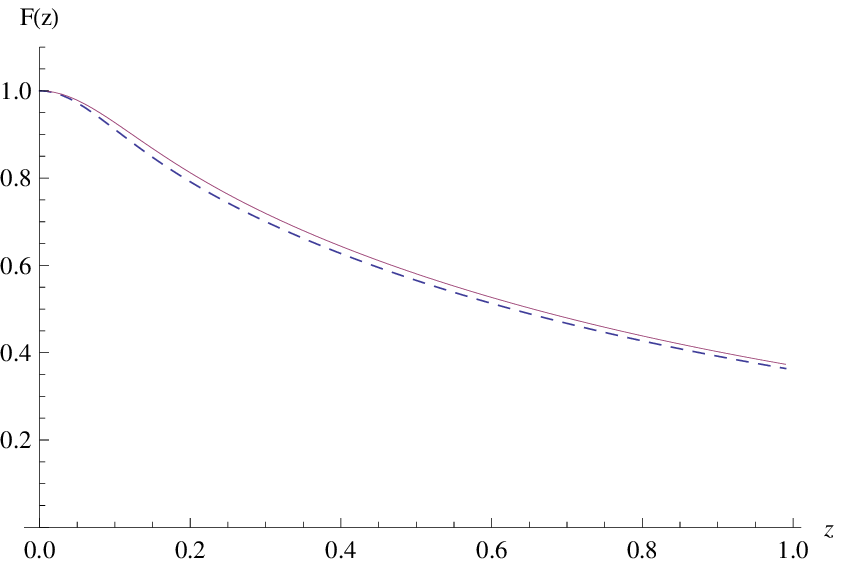}
\includegraphics[width=.3\textwidth]{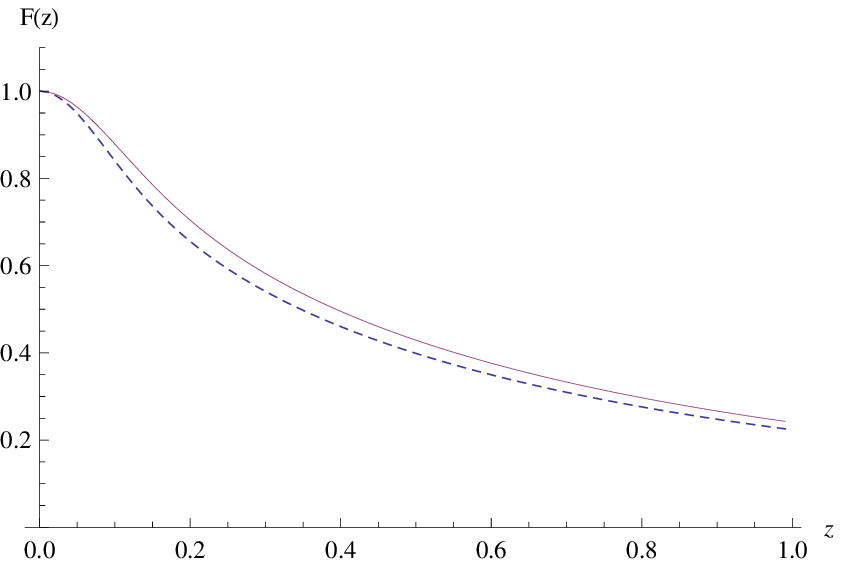}
\caption{The field $F$ (eq.\ (\ref{eq8})) for $\Delta = 1.6$ (left panel), $1.8$ (middle panel), $2$ (right panel) and $d=4$. Solid curves are first-order analytic expression (\ref{eq12F}), and dashed curves are exact numerical results (almost indistinguishable) at $T/T_c \approx 0.2$.}
\label{fig2a}
\end{figure}
Before we consider the next iterative order, we note that at finite temperature, the first-order expression (\ref{eq12F}) is in excellent agreement with numerical results even at $T/T_c \sim 0.1$, which is the lowest temperature at which a numerical solution is available. This is shown in figs.~\ref{fig2} and \ref{fig2a} in which the corresponding curves are almost indistinguishable, implying that the next iterative order introduces negligible corrections to the first-order expression (\ref{eq12F}) for temperatures $T/T_c \lesssim 0.1$.

We can repeat the above steps for the next iterative order to calculate $F_0^{(2)}$ and $\hat{\Phi}_0^{(2)}$. The resulting functions are very close to the their first-order counterparts, showing that the iteration converges rather rapidly.
In fact, the second order quantities are subleading in $1/b$. This is the case for all values of the scaling dimension $\Delta < \frac{d}{2}$.
\begin{figure}[t]
\includegraphics[width=.5\textwidth]{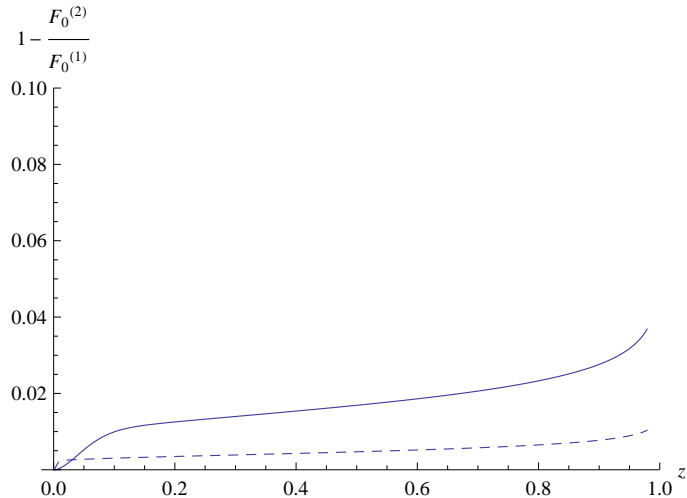}
\caption{The second order correction to the scalar field $F$ for $d=3$, $\Delta = 1.4$ at $b=20$ ($T/T_c \sim 0.1$) (solid line) and $b=200$ ($T/T_c \sim 0.01$) (dashed line).}
\label{figerr1}
\end{figure}
\begin{figure}[ht]
\includegraphics[width=.4\textwidth]{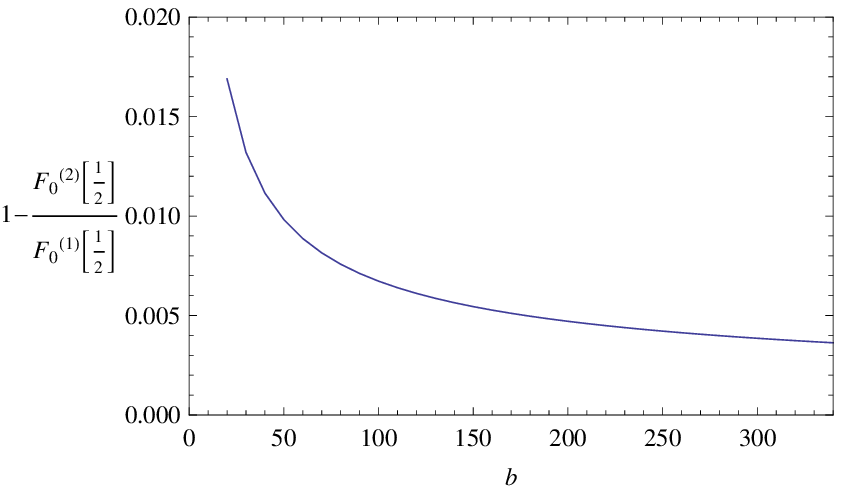}$\ \ \ \ $
\includegraphics[width=.4\textwidth]{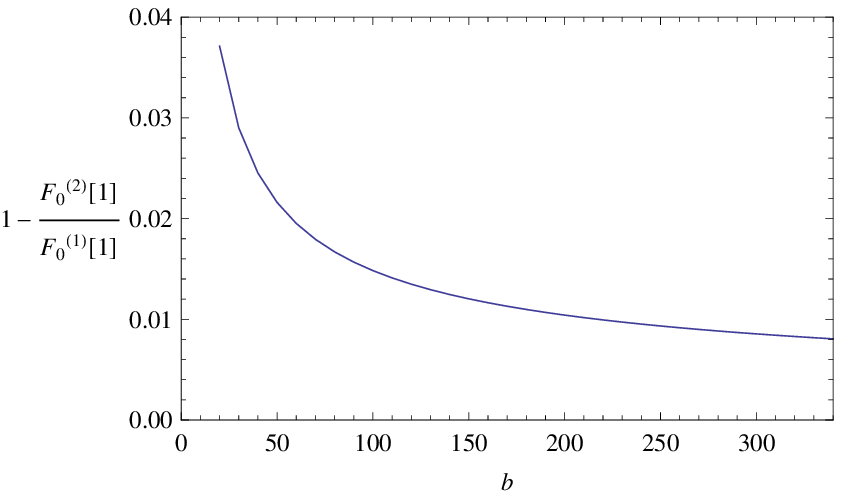}
\caption{The second order correction to the scalar field $F$ for $d=3$, $\Delta =1.4$ as a function of temperature at the mid-point, $z=\frac{1}{2}$, (left panel) and the horizon, $z=1$, (right panel). The horizontal axis corresponds to the temperature range $0.01 \lesssim \frac{T}{T_c} \lesssim 0.1$ with $T$ decreasing to the right.}
\label{figerr2}
\end{figure}
In fig.~\ref{figerr1}, we show the difference between first and second iterative order for $d=3$, $\Delta =1.4$ (all other values of $\Delta$ are similar). The error, $1-\frac{F_0^{(2)}}{F_0^{(1)}}$, is less than 5\% in the entire interval $[0,1]$.
As the temperature decreases from $T/T_c \sim 0.1$ to $T/T_c \sim 0.01$, the error decreases to less than $0.01$. To demonstrate that the error is subleading in $1/b$, in fig.~\ref{figerr2} we plot it at the mid-point ($z=\frac{1}{2}$) as well as the horizon ($z=1$) (at $z=0$ the error vanishes by design). Evidently, it goes to zero as $1/b$, showing that the second iterative order introduces subleading corrections at low temperature.

\begin{figure}[t]
\includegraphics[width=.5\textwidth]{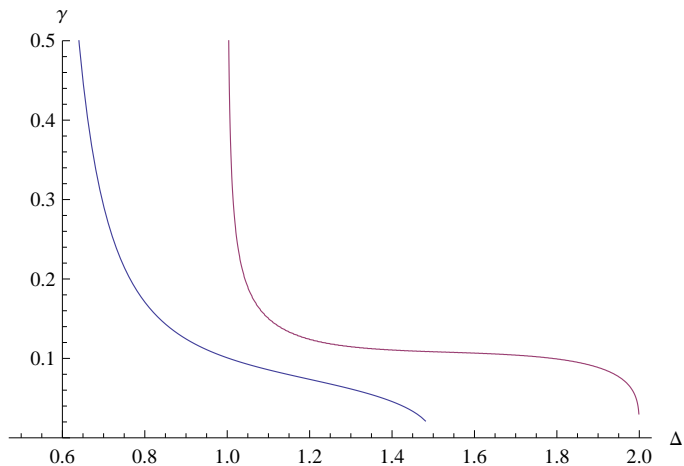}
\caption{The parameter $\gamma$ in the low temperature expression (\ref{eq30}) for the condensate {\em vs} $\Delta$. Curve on left (right) is for $d=3$ ($d=4$).}
\label{fig3}
\end{figure}
For the charge density we deduce from (\ref{eq15})
\be \rho_0 \sim b^{\frac{d}{2} + \Delta -1} \ee
Using
\be\label{eq29a} \frac{\langle q\mathcal{O}_\Delta\rangle^{1/\Delta}}{T_c} \sim b\rho_0^{-\frac{1}{d-1}} \ \ , \ \ \ \ \frac{T}{T_c} \sim \rho_0^{-\frac{1}{d-1}} \ee
we finally obtain for the energy gap
\be\label{eq30} E_g = \frac{\langle q\mathcal{O}_\Delta\rangle^{1/\Delta}}{T_c} = \gamma \left( \frac{T}{T_c} \right)^{-\frac{d/2-\Delta}{d/2+\Delta -1}} \ee
showing that the condensate diverges at low tempartures (to be precise, in the regime \eqref{eq21}). The exponent depends on the dimensions of the operator and spacetime. The expression for the exponent in (\ref{eq30}) corrects an earlier analytic result \cite{Siopsis:2010uq}.
The constant of proportionality $\gamma$ can be found analytically. It is plotted in fig.\ \ref{fig3} {\em vs} $\Delta$.
As $\Delta$ approaches the BF bound, $\gamma\to 0$, showing that the power law behavior changes, as we discuss next.

Indeed, at the end point (BF bound), $\Delta = \frac{d}{2}$, we need to exercise care. Letting $\Delta = \frac{d}{2} - \epsilon$, we obtain from (\ref{eqc1}) and (\ref{eqa1a2}),
\be \frac{1}{\mu^2} = \frac{(d-2)\Gamma(\frac{2}{d})}{d^{2(1-\frac{2}{d})}\Gamma(1-\frac{2}{d})} \left[ \frac{1}{2\epsilon b^{2-2\epsilon}} - \frac{1}{2\epsilon b^2} + \dots \right] \ee
and taking the limit $\epsilon\to 0$, we deduce at the BF bound
\be \frac{\mu^2}{b^2} = \frac{d^{2(1-\frac{2}{d})}\Gamma(1-\frac{2}{d})}{(d-2) \Gamma(\frac{2}{d})[ \ln b + \beta_d + o(b^0) ] } \ee
where $\beta_d$ is a constant that depends on the dimension and is easily computed (e.g., for $d=3$, $\beta_3 \approx 1.75$).

Higher-order corrections are computed as before by considering the next iterative order. This introduces corrections that are subleading as $b\to \infty$.
Already at $T/T_c \sim 0.1$, our first-order analytic results are almost indistinguishable from numerical results (see right panels of figs.~\ref{fig2} and \ref{fig2a}), the discrepancy being $\sim 1\%$. Below that temperature, numerical results are not available. Nevertheless the error can be estimated by calculating the correction at the second iterative order, as before. One finds that the error tends to zero as $b\to \infty$ in the entire interval $[0,1]$.

For the charge density, we have
\be \rho_0 \sim b^{d-1} (\ln b)^{-1/2} \ee
therefore the energy gap behaves as
\be\label{eqC1} E_g = \frac{\langle q\mathcal{O}_\Delta\rangle^{1/\Delta}}{T_c} \sim (\ln b)^{\frac{1}{2(d-1)}} \sim \left( \ln \frac{T_c}{T} \right)^{\frac{1}{2(d-1)}} \ee
showing that the condensate diverges at the BF bound, albeit very mildly.
This mild divergence was missed in earlier numerical studies \cite{Horowitz:2008bn}.

The BF bound can also be approached from above. However, the calculation becomes considerably more invloved, because for $\Delta > d/2$, as $T\to 0$, we have $F_0\approx 1$ near the boundary ($z=0$), but asymptotically ($z\gtrsim 1/b$),
$F_0 \sim z^{d-2\Delta}$,
which does not have a smooth limit as $T\to 0$. Therefore, we cannot apply perturbation theory and a different approach is called for \cite{Siopsis:2010uq}.
For example, one can approximate $F_0$ by
\be\label{eqFd} F_0(z) = \left\{ \begin{array}{ccc} 1 & , & z\le \alpha \\ \left( \frac{z}{\alpha} \right)^{d-2\Delta} & , & z>\alpha \end{array} \right. \ee
and find $\alpha$ by a variational method.
We shall not dwell on this further here.

Having understood the probe limit, we now turn to the first-order corections in a $1/q^2$ expansion.
For $\Delta < d/2$, it is necessary to include these corrections in order to obtain a physical system at low temperatures, because in the $q\to\infty$ limit the condensate diverges as $T\to 0$ (eqs.~(\ref{eq30}) and (\ref{eqC1})).

At first order, we obtain for the functions determining the metric,
\bea
zf_1' -df_1 &=& \frac{(bz)^{2\Delta}}{4(d-1)} \left[ \left( m^2 +\Delta^2 f_0 +
\frac{z^2 \Phi_0^2}{(bz)^{2\Delta}} \right) F_0^2 + 2\Delta z f_0 F_0 F_0' +
z^2 f_0 {F_0'}^2 + \frac{z^4}{(bz)^{2\Delta}} {\Phi_0'}^2 \right] \nonumber\\
z\chi_1' &=& \frac{(bz)^{2\Delta}}{d-1} \left[ \left( \Delta^2 + \frac{z^2\Phi_0^2}{f_0^2} \right) F_0^2 + 2\Delta z F_0 F_0' + z^2 {F_0'}^2 \right]
\eea
They can be solved at low temperature using our zeroth-order results above. We obtain
\be f_1(z) = -\frac{\Delta}{4(d-1)} (bz)^{2\Delta} \left[ 2 - z^d - z^{d-2\Delta} \right] + \dots \ , \ \
\chi_1(z) = - \frac{\Delta}{2(d-1)} (bz)^{2\Delta} + \dots \ee
For the temperature, we deduce the first-order expression
\be T = \frac{d}{4\pi} \left[ 1 + \frac{\Delta^2}{2d(d-1)} \frac{b^{2\Delta}}{q^2} + \dots \right] \ee
showing that the temperature receives a positive correction away from the probe limit.
Moreover, it is now clear when the probe limit fails. Indeed, for the expansion in $1/q^2$ to be valid, we ought to have
\be\label{eq50} b \lesssim q^{1/\Delta} \ee
justifying our earlier definition of the low energy regime \eqref{eq21}.

For a given $q$, \eqref{eq50} places a lower bound on the temperature. While zero temperature is unattainable for finite $q$, the temperature can be made arbitrarily low by choosing a sufficiently large $q$. It follows that, even though the probe limit ($q\to\infty$) is not a physical system at zero temperature, its properties are a good approximation to corresponding properties of physical systems (of finite charge $q$ and (low) temperature). The approximation becomes better with increasing $q$ as the $1/q^2$ corrections become smaller.

\section{Conductivity}
\label{sec3}

Next, we calculate the low temperature conductivity at the BF bound. For explicit analytic results, we concentrate on two cases, $d=3$ and $d=4$.
We shall obtain the conductivity $\sigma$ as a function of the rescaled frequency
\be \hat\omega = \frac{\omega}{b} = \frac{\omega}{\langle q\mathcal{O}_\Delta\rangle^{1/\Delta}} \ee
The function $\sigma (\hat\omega)$ has a well-defined limit as $q\to\infty$ (probe limit) down to low temperatures (in the regime \eqref{eq21}) at which the condensate $\langle\mathcal{O}_\Delta\rangle$ is large.
The conductivity of physical systems can be obtained as a $1/q^2$ expansion with the conductivity in the probe limit serving as the zeroth order term in the expansion.

\subsection{Three dimensions}

The conductivity on the AdS boundary is found by applying a sinusoidal electromagnetic perturbation in the bulk of frequency $\omega$ obeying the
wave equation
\be\label{eq53} - \frac{d^2 A}{dr_*^2} + V A = \omega^2 A \ \ , \ \ \ \ V = \frac{2q^2}{z^2} f\Psi^2 \ee
where $A$ is any component of the perturbing electromagnetic potential along the boundary.
Eq.~(\ref{eq53}) is to be solved subject to ingoing boundary condition at the horizon
\be\label{eq54} A \sim e^{-i\omega r_*} \sim (1-z)^{-i\omega/3} \ee
as $z\to 1$ ($r_*\to -\infty$), where $r_*$ is the tortoise coordinate
\be r_* = \int \frac{dz}{f(z)} = \frac{1}{6} \left[ \ln \frac{(1-z)^3}{1-z^3}
-2\sqrt 3 \tan^{-1} \frac{\sqrt 3 z}{2+z} \right] \ee
with the integration constant chosen so that the boundary is at $r_* =0$.
We with to solve this equation at low temperatures.

Using (\ref{eq8}) with $d=3$,
%For $\Delta \le 3/2$, we have
%\be \Psi = \frac{\langle \mathcal{O}_\Delta \rangle}{\sqrt{2} } z^{\Delta} F(z) \ee
the wave equation reads
\be\label{eq39} \frac{d}{dz} \left[ (1-z^3) \frac{dA}{dz} \right] - \left[ b^{2\Delta} z^{2\Delta -2} F^2(z) - \frac{\omega^2}{1-z^3} \right] A = 0
\ee
To account for the boundary condition at the horizon, set
\be A = (1-z)^{-i\omega/3} e^{-i\omega z/3} \mathcal{A} (z) \ee
where we included a factor $e^{-i\omega z/3}$ for convenience, so that only $\mathcal{A} (z)$ will contribute to the conductivity.
The wave equation becomes
\bea\label{eq44} & & - 3(1-z^3) \mathcal{A}'' + z\left[ 9z - 2(1+z+z^2)i\omega \right] \mathcal{A}'
\nonumber\\
& & + \left[ 3b^{2\Delta} z^{2\Delta -2} F^2(z) - (1+2z+3z^2)i\omega - \frac{(3+2z+z^2)(3+z+z^2+z^3)}{3(1+z+z^2)} \omega^2 \right] \mathcal{A} = 0 \eea
Regularity of the wavefunction $\mathcal{A}$ at the horizon ($z=1$) implies the boundary condition
\be\label{eq24} \left( 3 - 2i\omega \right) \mathcal{A}' (1) + \left( b^{2\Delta} F^2(1) - 2i\omega - \frac{4\omega^2}{3} \right) \mathcal{A} (1) = 0 \ee
At low temperatures, $b\gg 1$, it is convenient to rescale $z\to z/b$. The wave equation can be solved as a series expansion in $1/q^2$.
The zeroth order term is given by replacing $F$ by $F_0$ (eq.~(\ref{eq5a})).
For $\Delta \le \frac{3}{2}$, $F_0(z)$ has a smooth limit (albeit unphysical) as $b\to\infty$ (eq.\ \eqref{eqFT0}).
After rescaling and letting $b\to\infty$, we obtain the wave equation
\be\label{eq44zero} - \mathcal{A}''
+ \left[ z^{2\Delta -2} - \hat{\omega}^2 \right] \mathcal{A} = 0
\ee
where we used $F(z/b) \to F(0)=1$, as $b\to\infty$.
For $1<\Delta \le \frac{3}{2}$, there are two linearly independent solutions, $\mathcal{A}_\pm$, distinguished by their asymptotic behavior,
\be \mathcal{A}_\pm \sim e^{\pm \frac{1}{\Delta} z^\Delta} \ \ , \ \ \ \ z\to\infty \ee
The general solution can be written as a linear combination,
\be \mathcal{A} = c^+ \mathcal{A}_+ + c^- \mathcal{A}_- \ee
Applying the boundary condition (\ref{eq24}), we deduce
\be \frac{c^+}{c^-} \sim e^{-\frac{2}{\Delta} b^{\Delta}} \ee
so at very low temperatures,
\be c^+ = 0 \ee
i.e., $\mathcal{A} \to 0$ as $z\to\infty$.

For $\Delta = \frac{3}{2}$, we obtain the exact explicit limit solution
\be \mathcal{A} (z) = \mathcal{A}_-(z) = \mathrm{Ai} (bz - \hat{\omega}^2 ) \ee
whereas $\mathcal{A}_+ (z) = \mathrm{Bi} (bz - \hat{\omega}^2 ) $,
with arbitrary normalization, where we restored the scaling parameter $b$.

In this (unphysical) limit, the quasinormal frequencies have moved to the real axis yielding an infinite set of normal frequencies which are solutions of
\be\label{eq47} \mathrm{Ai} (-\hat{\omega}^2) = 0 \ee
Thus we obtain an infinite tower of real frequencies given by the zeroes of the Airy function.

We deduce the limit conductivity (as $q,b\to\infty$)
\be\label{eq55} \sigma (\hat\omega) = \frac{i}{\hat\omega} \frac{\mathrm{Ai}' (-\hat{\omega}^2) }{\mathrm{Ai} (-\hat{\omega}^2)
} \ee
The real frequencies that solve (\ref{eq47}) are the poles of the conductivity. Notice that $\Re\sigma = 0$, except at the poles of $\Im\sigma$ where $\Re\sigma$ diverges as a $\delta$-function.

This expression is unphysical, as we have already explained (the probe limit breaks down in the limit $b\to\infty$), but it is useful for computational purposes. We shall use it as a starting point to calculate the conductivity of a physical system at low temperatures.

At low temperatures, we can calculate the first-order correction analytically
by considering the $b\to\infty$ wave equation (\ref{eq44zero}) as the zeroth-order equation. Then for the first-order correction $\delta\mathcal{A}$ to the potential at low temperatures, we obtain from (\ref{eq44}),
\be\label{eq44f} -\delta\mathcal{A}'' + [z-\hat\omega^2 ] \delta\mathcal{A} = - \frac{1}{3(1-z^3)} \mathcal{H}_1 \mathcal{A} \ee
where
\be \mathcal{H}_1 = z\left[ 9z - 2(1+z+z^2)i\omega \right] \frac{d}{dz}
+ 3b^{3} z (2F_1(z) +z^3) - (1+2z+3z^2)i\omega + \frac{z^2(1-15z-12z^2-10z^3)}{3(1+z+z^2)} \omega^2 \ee
The first-order potential leads to quasinormal modes which are zeroes of $\mathcal{A} + \delta\mathcal{A}$.
Thus the real frequencies (\ref{eq47}) of the (unphysical) $b\to\infty$ limit
get shifted at low temperatures away from the real axis. We obtain $\hat\omega \to \hat\omega + \delta\hat\omega$, where
\be\label{eqdo} \delta\hat{\omega} =  \frac{\pi \mathrm{Bi} (-\hat{\omega}^2)}{3\hat{\omega} \mathrm{Ai}' (-\hat{\omega}^2)} \int_0^1 \frac{dz}{1-z^3}\, \mathrm{Ai} (bz - \hat\omega^2) \mathcal{H}_1 \mathrm{Ai} (bz-\hat{\omega}^2) \ee
This first-order expression is valid for low frequencies. As we heat up the system, most modes disappear and we are left with a finite number of quasinormal modes. Their number decreases as we increase the temperature.
Conversely, as we cool down the system, (\ref{eqdo}) becomes increasingly accurate for an increasing number of modes. These modes shift toward the real axis ($\delta\hat\omega \to 0$) as we lower the temperature.
In the low temperature regime \eqref{eq21}, the large $q$ is, the more spikes one obtains for $\sigma (\omega)$.

\begin{figure}
\includegraphics{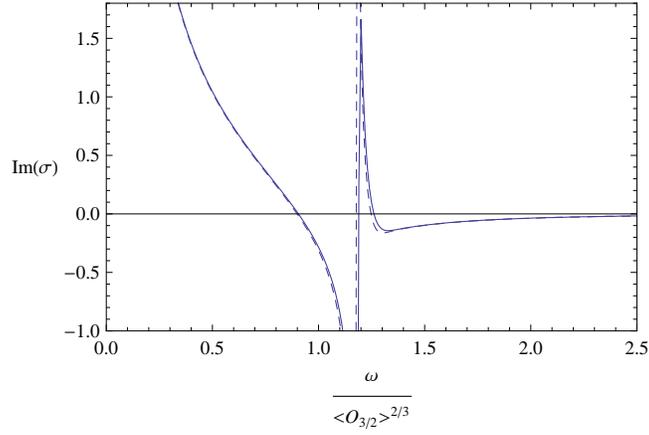}
\caption{The imaginary part of the conductivity in $d=3$ using the expression (\ref{eq12F}) for the scalar field (dotted line) compared with the exact numerical solution (solid line) at $ \frac{T}{T_c}\approx.1$}
\label{fig01}
\end{figure}

\begin{figure}[ht]
\includegraphics{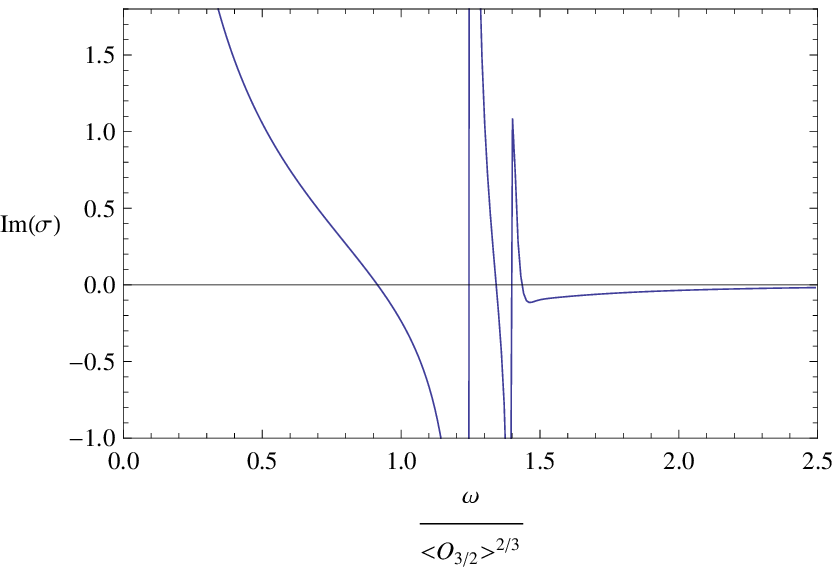}
\includegraphics{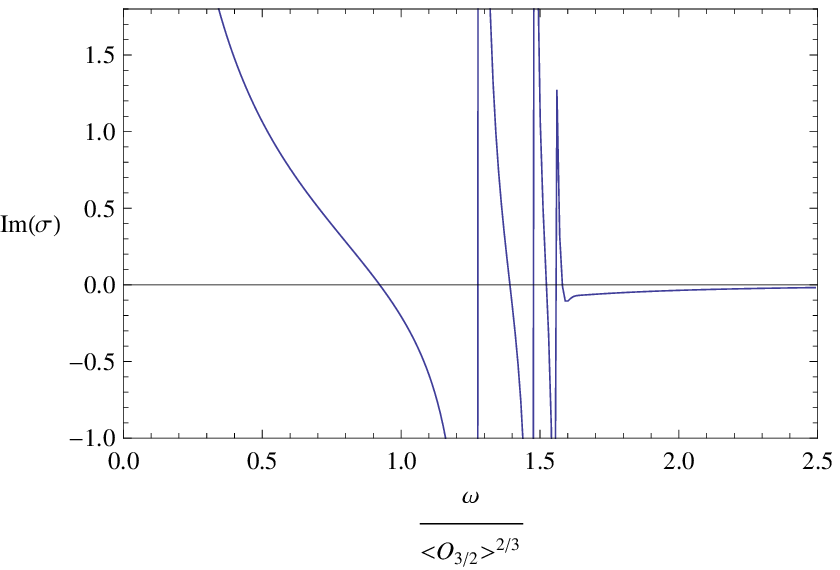}
\caption{The imaginary part of the conductivity {\em vs.}~frequency in $d=3$ using the expression (\ref{eq12F}) for $F$ at $\frac{T}{T_c}\approx .05$ (left), $.04$ (right). As the temperature decreases, poles move toward the real axis.}
\label{figsigintermediate}
\end{figure}

\begin{figure}[ht]
\includegraphics{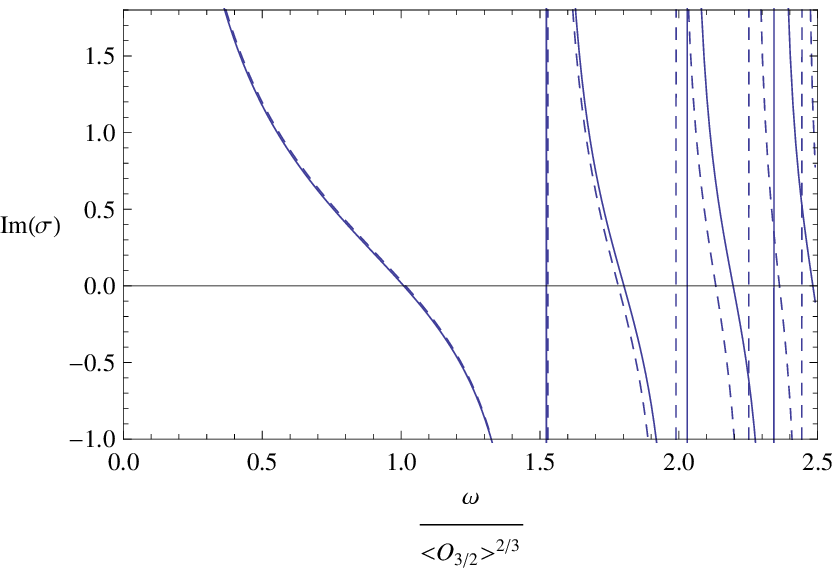}
\caption{Comparison of the imaginary part of the conductivity in $d=3$ using the expression (\ref{eq12F}) for F at $\frac{T}{T_c}\approx .01$ (dotted line) and the $b\to\infty$ limit (\ref{eq55}) (solid line).}
\label{figsigavsap}
\end{figure}

This shifting of quasinormal modes can be seen in plots of the conductivity.
As the mode frequency approaches the real axis, the corresponding spike in the plot of the imaginary part of the conductivity becomes more pronounced. 
To demonstrate this, we calculated the conductivity using the first-order approximation (\ref{eq12F}) to the scalar field.
In figure \ref{fig01}, we show the imaginary part of the conductivity at temperature $T/T_c \approx .1$ and compare with the exact numerical solution. The agreement is very good even at such high temperature at which only one quasinormal mode is left.
Unfortunately, this is the low temperature limit attained by numerical analysis as numerical instabilities prohibit one from lowering the temperature further.
Using our analytic results, we see in figure \ref{figsigintermediate} the emergence of an increasing number of poles as we lower the temperature to $T/T_c \approx .06$ and $.04$.
Finally in figure \ref{figsigavsap} we compare the lower temperature ($T/T_c\approx .01$) result with the $b\to\infty$ limit analytic expression (\ref{eq55}), thus demonstrating convergence.


For $\Delta > 3/2$, the potential is
\be V = b^{2\Delta} z^{2\Delta -2} (1-z^3) F(bz) \ee
with $F$ given approximately by (\ref{eqFd}).
It attains a maximum of order $b^{2(2-\Delta)}$ for $\Delta < 2$. Therefore, at zero temperature it has infinite height. However, the width becomes infinitely narrow leading to a finite tower of poles for the conductivity (quasinormal modes).
In the zero temperature limit, the number of modes increases as one approaches the BF bound and decreases away from it.
For $\Delta \ge 2$, the height of the potential becomes finite at zero temperature.
It turns out that the potential is too narrow to possess bound states, so no poles exist for $\Delta \ge 2$.

\subsection{Four dimensions}

The $d=4$ case is similar.
Working as in the $d=3$ case, in the (unphysical) $b\to\infty$ limit, the wave equation for $\Delta = 2$ (at the BF bound) in the probe limit reduces to
\be A'' - \frac{1}{z} A' - [ b^4 z^2 - \omega^2 ] A = 0 \ee
whose acceptable solution can be written in terms of a Whittaker function,
\be A = W_{\frac{\hat\omega^2}{4} , \frac{1}{2}} ( b^2z^2 ) \ee
(The other solution diverges as $z\to\infty$.)
At the boundary ($z\to 0$), it has a logarithmic divergence which we need to subtract before we can calculate QNMs and the conductivity \cite{Horowitz:2008bn}. The conductivity is then given by
\be
\sigma(\hat\omega)=\frac{2}{i\hat\omega} \frac{A_2}{A_0}+\frac{i\hat\omega}{2}
\ee
where
\be A (z) = A_0 + A_2 b^2z^2 - A_0 \frac{\hat\omega^2}{2} b^2z^2\ln (b^2z^2) + \dots \ee
with an arbitrarily chosen cutoff.

Using the expansion for small arguments,
\be  W_{\frac{\hat\omega^2}{4} , \frac{1}{2}} ( b^2z^2 ) = - \frac{2}{\hat\omega^2 \Gamma (-\hat\omega^2/4)} \left\{ 1 - \left[ 1 + \hat\omega^2 \left( 2\gamma -1 + \ln (b^2z^2) + \psi (1-\hat\omega^2/4) \right) \right] \frac{b^2 z^2}{2} \right\} + \dots \ee
we deduce the limit conductivity as $q, b\to\infty$,
\be\label{eq55d4} \sigma(\hat\omega) = \frac{1}{i\hat\omega} + i\hat\omega \left[ 2\gamma - \frac{1}{2} + \psi (1- \hat\omega^2/4) \right] \ee
We have a pole at $\omega = 0$, as expected and an infinite tower of real poles determined by the poles of the digamma function. The poles have real frequencies
\be\label{eqC2} \hat\omega = \frac{\omega_n}{\langle\mathcal{O}\rangle^{1/2}} = 2\sqrt{n} \ \ , \ \ \ \ n=0,1,2,\dots \ee
\begin{figure}[t]
\includegraphics{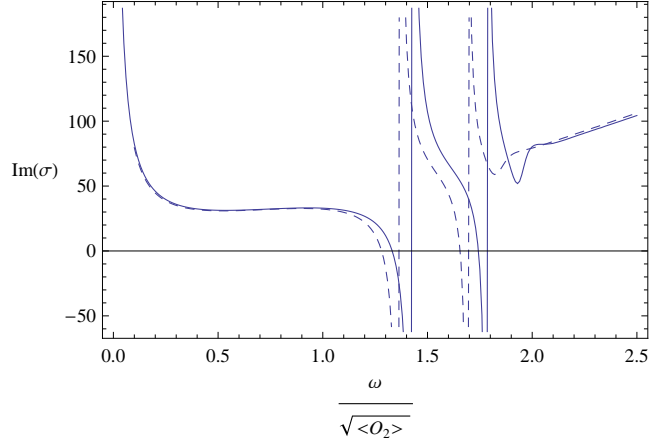}
\caption{The imaginary part of the conductivity in $d=4$ using the expression (\ref{eq12F}) for the scalar field (dotted line) compared with the exact numerical solution (solid line) at $ \frac{T}{T_c}\approx.17$}
\label{fig01d4}
\end{figure}
These poles are unphysical, but are good approximations to the quasinormal modes at low temperatures, in the physical regime \eqref{eq21}. As we increase the temperature, these modes move away from the real axis. At any given temperature we have a finite number of such modes. As we lower the temperature, their number increases and the quasinormal modes approach the poles \eqref{eqC2} on the real axis.
To demonstrate this, we have calculated the conductivity using the first-order approximation (\ref{eq12F}) to the scalar field.
In figure \ref{fig01d4} we compare with numerical results at temperature $T/T_c \approx .17$ and find good agreement.
As we go to lower temperature, numerical instabilities arise and it is no longer possible to compare our analytic results with their numerical counterparts.
We find convergence to the $b\to\infty$ limit (\ref{eq55d4}), but much slower than in $d=3$. In figure
\ref{figsigintermediated4} we show the imaginary part of the conductivity at $T/T_c  \approx 0.1$ and $0.04$. As we lower the temperature, the number of poles increases and the poles shift to the right on the real axis approaching the limit values (\ref{eqC2}) which correspond to the $b\to\infty$ limit of the conductivity shown in figure \ref{figsigavsapd4}.
\begin{figure}
\includegraphics{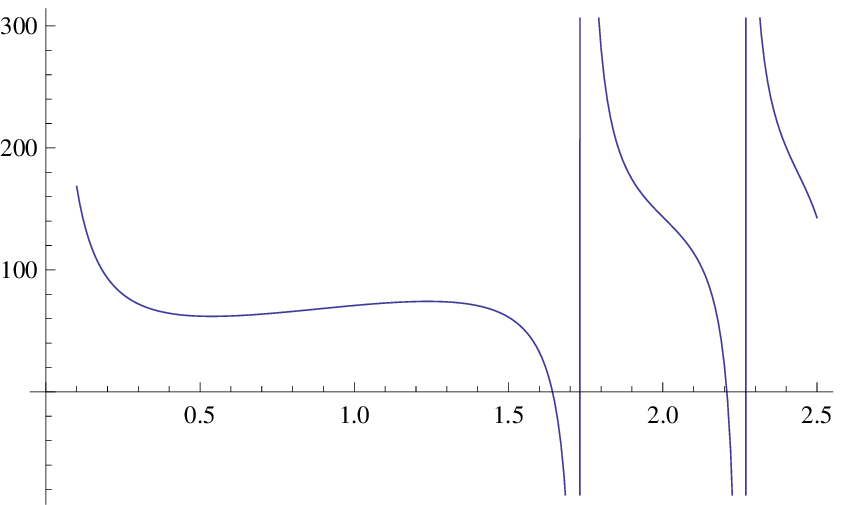}
\includegraphics{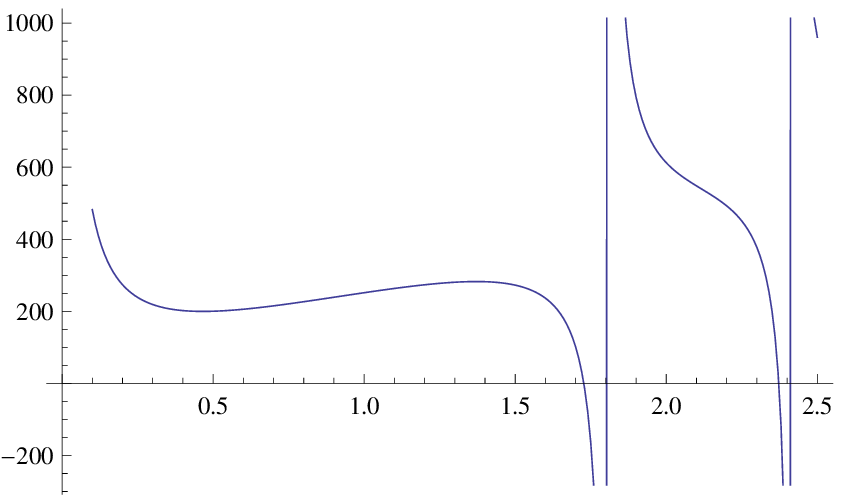}
\caption{The imaginary part of the conductivity {\em vs.}~frequency in $d=4$ using the expression (\ref{eq12F}) for $F$ at $\frac{T}{T_c}\approx .1$ (left), $.04$ (right).}
\label{figsigintermediated4}
\end{figure}
\begin{figure}[ht]
\includegraphics{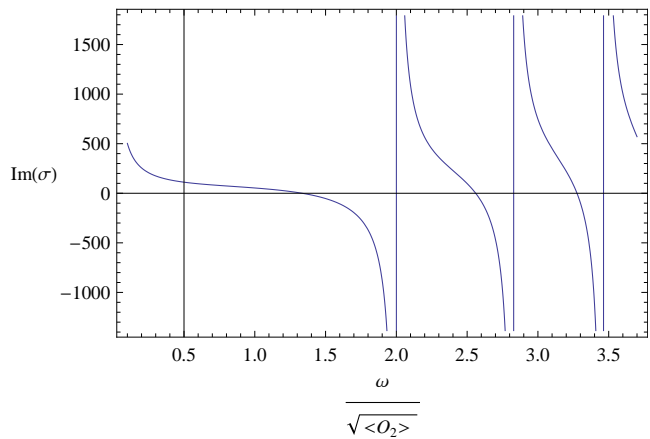}
\caption{The imaginary part of the conductivity in the (unphysical) limit $b\to\infty$ in $d=4$ (\ref{eq55d4}).}
\label{figsigavsapd4}
\end{figure}

\section{Conclusion}
\label{sec4}

We discussed holographic superconductors in the probe limit when the scalar hair of the dual black hole is near the BF bound.
Using the analytic tools developed in \cite{Siopsis:2010uq}, we analyzed the low temperature regime trying to go beyond the range of applicability of numerical techniques which fail at very low temperatures.
Thus undeterred by numerical instabilities, we found that at low temperatures the condensate diverges as $|\ln T|^\delta$ where $\delta$ depends on the dimension of spacetime (eq.~(\ref{eqC1})).
This signals the breakdown of the probe limit at low temperatures even at the BF bound.
The divergence is very mild which explains why it was missed in earlier numerical analyses.
Even though the probe limit at zero temperature cannot be a physical state for scaling dimensions $\Delta \le \frac{d}{2}$, it is still useful to analyze it, because it is a good approximation to physical states at low temperatures and thus a good starting point for computational purposes.

We calculated the low temperature conductivity at the BF found in the probe limit and found exact limit analytic expressions in $d=3,4$. Thus, we showed that as we lower the temperature, the conductivity has an increasing number of spikes (quasinormal modes) which approach the infinite tower of real poles determined by the zeroes of the Airy function in $d=3$ (eq.~(\ref{eq47})) and the Gamma function in $d=4$ (eq.~(\ref{eqC2})).
As we heat up the system, the number of spikes decreases and their positions move off of the real axis.

The probe limit we studied here can be used as a zeroth-order contribution to a perturbative expansion in $1/q^2$.
Our results can be extended in a systematic way to include back reaction to the bulk metric in order to analyze physical states. This will greatly facilitate the probe of the zero temperature limit and shed light on the ground state of the system, as numerical methods fail due to numerical instabilities. Work in this direction is in progress.

\section*{Acknowledgment}

%S.~M.~was partially supported by the Thailand Research Fund.
%G.~S.~and J.~T.~were 
Work supported in part by the Department of Energy under grant DE-FG05-91ER40627.
S.~M.~gratefully acknowledges the hospitality of the Department of Physics and
Astronomy at the University of Tennessee where part of the work was performed.


\begin{thebibliography}{99}


 %\cite{Maldacena:1997re}
\bibitem{Maldacena:1997re}
  J.~M.~Maldacena,
  %``The large N limit of superconformal field theories and supergravity,''
  Adv.\ Theor.\ Math.\ Phys.\  {\bf 2}, 231 (1998)
  [Int.\ J.\ Theor.\ Phys.\  {\bf 38}, 1113 (1999)]
  [arXiv:hep-th/9711200].
  %%CITATION = IJTPB,38,1113;%% 
 
%\cite{Gubser:2008px}
\bibitem{Gubser:2008px}
  S.~S.~Gubser,
  %``Breaking an Abelian gauge symmetry near a black hole horizon,''
  Phys.\ Rev.\  D {\bf 78}, 065034 (2008)
  [arXiv:0801.2977 [hep-th]].
  %%CITATION = PHRVA,D78,065034;%%

 %\cite{Hartnoll:2008vx}
\bibitem{Hartnoll:2008vx}
  S.~A.~Hartnoll, C.~P.~Herzog and G.~T.~Horowitz,
  %``Building a Holographic Superconductor,''
  Phys.\ Rev.\ Lett.\  {\bf 101}, 031601 (2008)
  [arXiv:0803.3295 [hep-th]].
  %%CITATION = PRLTA,101,031601;%% 
  
  %\cite{Horowitz:2008bn}
\bibitem{Horowitz:2008bn}
  G.~T.~Horowitz and M.~M.~Roberts,
  %``Holographic Superconductors with Various Condensates,''
  Phys.\ Rev.\  D {\bf 78}, 126008 (2008)
  [arXiv:0810.1077 [hep-th]].
  %%CITATION = PHRVA,D78,126008;%%
  
%\cite{Franco:2009yz}
\bibitem{Franco:2009yz}
  S.~Franco, A.~Garcia-Garcia and D.~Rodriguez-Gomez,
  %``A general class of holographic superconductors,''
  arXiv:0906.1214 [hep-th].
  %%CITATION = ARXIV:0906.1214;%%

%%\cite{Albash:2008eh}
%\bibitem{Albash:2008eh}
%  T.~Albash and C.~V.~Johnson,
%  %``A Holographic Superconductor in an External Magnetic Field,''
%  JHEP {\bf 0809}, 121 (2008)
%  [arXiv:0804.3466 [hep-th]].
%  %%CITATION = JHEPA,0809,121;%%
%  
%%\cite{Hartnoll:2008kx}
%\bibitem{Hartnoll:2008kx}
%  S.~A.~Hartnoll, C.~P.~Herzog and G.~T.~Horowitz,
%  %``Holographic Superconductors,''
%  JHEP {\bf 0812}, 015 (2008)
%  [arXiv:0810.1563 [hep-th]].
%  %%CITATION = JHEPA,0812,015;%%  
 
 %\cite{Hartnoll:2007ai}
\bibitem{Hartnoll:2007ai}
  S.~A.~Hartnoll and P.~Kovtun,
%  ``Hall conductivity from dyonic black holes,''
  Phys.\ Rev.\  D {\bf 76}, 066001 (2007)
  [arXiv:0704.1160 [hep-th]].
  %%CITATION = PHRVA,D76,066001;%%

%\cite{Hartnoll:2007ih}
\bibitem{Hartnoll:2007ih}
  S.~A.~Hartnoll, P.~K.~Kovtun, M.~Muller and S.~Sachdev,
%  ``Theory of the Nernst effect near quantum phase transitions in condensed
%  matter, and in dyonic black holes,''
  Phys.\ Rev.\  B {\bf 76}, 144502 (2007)
  [arXiv:0706.3215 [cond-mat.str-el]].
  %%CITATION = PHRVA,B76,144502;%%

%\cite{Hartnoll:2007ip}
\bibitem{Hartnoll:2007ip}
  S.~A.~Hartnoll and C.~P.~Herzog,
%  ``Ohm's Law at strong coupling: S duality and the cyclotron resonance,''
  Phys.\ Rev.\  D {\bf 76}, 106012 (2007)
  [arXiv:0706.3228 [hep-th]].
  %%CITATION = PHRVA,D76,106012;%%

%\cite{Hartnoll:2008hs}
\bibitem{Hartnoll:2008hs}
  S.~A.~Hartnoll and C.~P.~Herzog,
%  ``Impure AdS/CFT,''
Phys.\ Rev.\ D {\bf 77}, 106009 (2008)
  [arXiv:0801.1693 [hep-th]].
  %%CITATION = ARXIV:0801.1693;%%

 
 %\cite{Hartnoll:2009sz}
\bibitem{Hartnoll:2009sz}
  S.~A.~Hartnoll,
  %``Lectures on holographic methods for condensed matter physics,''
  Class.\ Quant.\ Grav.\  {\bf 26}, 224002 (2009)
  [arXiv:0903.3246 [hep-th]].
  %%CITATION = CQGRD,26,224002;%% 
  
  %\cite{Herzog:2009xv}
\bibitem{Herzog:2009xv}
  C.~P.~Herzog,
  %``Lectures on Holographic Superfluidity and Superconductivity,''
  J.\ Phys.\ A  {\bf 42}, 343001 (2009)
  [arXiv:0904.1975 [hep-th]].
  %%CITATION = JPAGB,A42,343001;%%
  
  %\cite{Horowitz:2010gk}
\bibitem{Horowitz:2010gk}
  G.~T.~Horowitz,
  %``Introduction to Holographic Superconductors,''
  arXiv:1002.1722 [hep-th].
  %%CITATION = ARXIV:1002.1722;%%
  
%\cite{Konoplya:2009hv}
\bibitem{Konoplya:2009hv}
  R.~A.~Konoplya and A.~Zhidenko,
  %``Holographic conductivity of zero temperature superconductors,''
  Phys.\ Lett.\  B {\bf 686}, 199 (2010)
  [arXiv:0909.2138 [hep-th]].
  %%CITATION = PHLTA,B686,199;%%
  
%\cite{Horowitz:2009ij}
\bibitem{Horowitz:2009ij}
  G.~T.~Horowitz and M.~M.~Roberts,
  %``Zero Temperature Limit of Holographic Superconductors,''
  JHEP {\bf 0911}, 015 (2009)
  [arXiv:0908.3677 [hep-th]].
  %%CITATION = JHEPA,0911,015;%%
  
  %\cite{Siopsis:2010uq}
\bibitem{Siopsis:2010uq}
  G.~Siopsis and J.~Therrien,
  %``Analytic calculation of properties of holographic superconductors,''
  JHEP {\bf 1005}, 013 (2010)
  [arXiv:1003.4275 [hep-th]].
  %%CITATION = JHEPA,1005,013;%%
  
   %\cite{Gubser:2009cg}
\bibitem{Gubser:2009cg}
  S.~S.~Gubser and A.~Nellore,
  %``Ground states of holographic superconductors,''
  Phys.\ Rev.\  D {\bf 80}, 105007 (2009)
  [arXiv:0908.1972 [hep-th]].
  %%CITATION = PHRVA,D80,105007;%%
  
%\cite{Faulkner:2009am}
\bibitem{Faulkner:2009am}
  T.~Faulkner, G.~T.~Horowitz, J.~McGreevy, M.~M.~Roberts and D.~Vegh,
  %``Photoemission 'experiments' on holographic superconductors,''
  arXiv:0911.3402 [hep-th].
  %%CITATION = ARXIV:0911.3402;%%

 
%\cite{Breitenlohner:1982jf}
\bibitem{Breitenlohner:1982jf}
  P.~Breitenlohner and D.~Z.~Freedman,
  %``Stability In Gauged Extended Supergravity,''
  Annals Phys.\  {\bf 144}, 249 (1982).
  %%CITATION = APNYA,144,249;%%
  
  \bibitem{M-T}  L.~Mezincescu and P.~K.~Townsend, Annals Phys.\ \textbf{160},
406 (1985). %%CITATION = APNYA,160,406;%%

  %\cite{Hertog:2004bb}
\bibitem{Hertog:2004bb}
  T.~Hertog and K.~Maeda,
  ``Stability and thermodynamics of AdS black holes with scalar hair,''
  Phys.\ Rev.\  D {\bf 71}, 024001 (2005)
  [arXiv:hep-th/0409314].
  %%CITATION = PHRVA,D71,024001;%%

  \bibitem{papa1}
  %\cite{Koutsoumbas:2009pa}
%\bibitem{Koutsoumbas:2009pa}
  G.~Koutsoumbas, E.~Papantonopoulos and G.~Siopsis,
%  ``Exact Gravity Dual of a Gapless Superconductor,''
  JHEP {\bf 0907}, 026 (2009)
  [arXiv:0902.0733 [hep-th]].
  %%CITATION = JHEPA,0907,026;%%

%\cite{Gregory:2009fj}
\bibitem{Gregory:2009fj}
  R.~Gregory, S.~Kanno and J.~Soda,
  %``Holographic Superconductors with Higher Curvature Corrections,''
  JHEP {\bf 0910}, 010 (2009)
  [arXiv:0907.3203 [hep-th]].
  %%CITATION = JHEPA,0910,010;%%



%\cite{Pan:2009xa}
\bibitem{Pan:2009xa}
  Q.~Pan, B.~Wang, E.~Papantonopoulos, J.~Oliveira and A.~Pavan,
  %``Holographic Superconductors with various condensates in
  %Einstein-Gauss-Bonnet gravity,''
  arXiv:0912.2475 [hep-th].
  %%CITATION = ARXIV:0912.2475;%%






\end{thebibliography}
\end{document}